\documentclass[amssymb,prb,twocolumn,floats,amsmath,showpacs,superscriptaddress]{revtex4}
\usepackage[T1]{fontenc}
\usepackage{bm}
\usepackage{graphicx}
\usepackage{amssymb}
\usepackage{amsfonts}
\usepackage{amsmath}

\begin{document}

\title{Effects of $s,p-d$ and $s-p$ exchange interactions probed by exciton magnetospectroscopy in (Ga,Mn)N}

\author{J. Suf\mbox{}fczy\'{n}ski}
\affiliation{Institute of Experimental Physics, University of Warsaw, ul. Ho\.za 69, PL-00-681 Warszawa, Poland}
\author{A. Grois}
\affiliation{Institut f\"ur Halbleiter- und Festk\"orperphysik, Johannes Kepler University, Altenbergerstr. 69, A-4040 Linz, Austria}
\author{W. Pacuski}\affiliation{Institute of Experimental Physics, University of Warsaw, ul. Ho\.za 69, PL-00-681 Warszawa, Poland}
\author{A. Golnik} \affiliation{Institute of Experimental Physics, University of Warsaw, ul. Ho\.za 69, PL-00-681 Warszawa, Poland}
\author{J. A. Gaj} \affiliation{Institute of Experimental Physics, University of Warsaw, ul. Ho\.za 69, PL-00-681 Warszawa, Poland}
\author{A. Navarro-Quezada}
\affiliation{Institut f\"ur Halbleiter- und Festk\"orperphysik, Johannes Kepler University, Altenbergerstr. 69, A-4040 Linz, Austria}
\author{B. Faina} \affiliation{Institut f\"ur Halbleiter- und Festk\"orperphysik, Johannes Kepler University, Altenbergerstr. 69, A-4040 Linz, Austria}
\author{T. Devillers} \affiliation{Institut f\"ur Halbleiter- und Festk\"orperphysik, Johannes Kepler University, Altenbergerstr. 69, A-4040 Linz, Austria}
\author{A. Bonanni} \affiliation{Institut f\"ur Halbleiter- und Festk\"orperphysik, Johannes Kepler University, Altenbergerstr. 69, A-4040 Linz, Austria}

\date{\today}


\begin{abstract}
Near band-gap photoluminescence and reflectivity in magnetic field are employed to determine the exchange-induced splitting of free exciton states in paramagnetic wurtzite Ga$_{1-x}$Mn$_x$N, $x \lesssim 1$\%, grown on sapphire substrates by metal-organic vapor phase epitaxy. The band gap is found to increase with $x$. The giant Zeeman splitting of all three excitons $A$, $B$ and $C$ is resolved, enabling the determination of the \textit{apparent} exchange integrals $N_{0}\alpha^{\text{(app)}} = 0.0 \pm 0.1$~eV and $N_0\beta^{(app)} = + 0.8 \pm 0.2$~eV. These non-standard values and signs of the  $s-d$ and $p-d$ exchange energies are explained in terms of recent theories that suggest a contribution of the electron-hole exchange to the spin splitting of the conduction band and a renormalization of the free hole spin-splitting by a large $p-d$ hybridization. According to these models, in the limit of a strong $p-d$ coupling, the band gap of (Ga,Mn)N increases with $x$ and the order of hole spin subbands is reversed, as observed.
\end{abstract}


%
\maketitle


\section{Introduction}

The quest for new spintronic functionalities stimulates the search for carrier-induced ferromagnetism in various families of dilute magnetic semiconductors (DMSs).\cite{Dietl:2000_S} Particular attention has been directed towards wide band gap semiconductors (primarily oxides\cite{Liu:2005_JMSME} and nitrides\cite{Bonanni:2007_SST}) containing magnetic ions, in view of the strong carrier-ion $p-d$ exchange coupling expected for these materials -- a prerequisite for room temperature ferromagnetism.\cite{Dietl:2000_S} However, it becomes increasingly clear that a further progress in this field requires a deeper understanding of these systems, particularly by studying magnetooptical phenomena that provide quantitative information on the dominant spin-dependent interactions. Interestingly, the recent research in this direction has revealed that the strong $p-d$ coupling affects the magnetooptical behavior in a surprising way,\cite{Dietl:2008_PRB} unanticipated  within the virtual crystal and the molecular-field approximations employed successfully over several decades for the description of the giant splittings of bands in moderate gap II-VI DMSs. Furthermore, it has been found that a meaningful description of magnetooptical phenomena, in addition to the $s,p-d$ exchange couplings, should take into account electron-hole exchange interactions within excitons\cite{Pacuski:2006_PRB} as well as between electrons residing in the conduction band and holes localized on magnetic ions.\cite{Sliwa:2008_PRB}

A reverse order of the exciton spin levels in the magnetic field, suggestive of the strong coupling limit of the $p-d$ interaction,\cite{Dietl:2008_PRB} was found for (Zn,Co)O,\cite{Pacuski:2006_PRB} (Ga,Mn)N,\cite{Marcet:2006_PRB} and (Ga,Fe)N.\cite{Pacuski:2008_PRL} In these systems, the transition metal
(TM) dopants act as isoelectronic impurities. Previous magnetooptical studies of (Ga,Mn)N near the fundamental absorption edge have allowed to evaluate the sum of \textit{apparent} exchange energies $N_{0}\alpha^{\text{(app)}}$ and $N_{0}\beta^{\text{(app)}}$ which parameterize the net giant Zeeman splittings of the conduction and valence band, respectively.\cite{Pacuski:2007_PRB}

In this work we present results of photoluminescence (PL) and reflectivity studies carried out as a function of temperature and magnetic field on Ga$_{1-x}$Mn$_x$N with Mn concentrations $x \lesssim 1$\%.  Our samples, obtained by metal-organic vapor phase epitaxy (MOVPE),  have been characterized by transmission electron microscopy (TEM), secondary-ion mass spectroscopy (SIMS), synchrotron x-ray diffraction (SXRD), extended x-ray absorption fine structure (EXAFS), and superconducting quantum interference device (SQUID) magnetometry.\cite{Stefanowicz:2010_PRBa} This palette of methods shows that the Mn ions are randomly distributed over cation sites and also that the predominant Mn charge state is 3+, implying a minimal compensation of Mn acceptors by residual donors.

The high quality of the samples studied here makes it possible to resolve all three fundamental excitons $A$, $B$, and $C$ specific to the wurtzite semiconductors, as well as to trace their shifts and splittings as a function of the Mn concentration $x$ and of the applied magnetic field $B$. While previous optical studies left the actual dependence of the (Ga,Mn)N band gap $E_{\text{g}}$ on $x$ still unsettled,\cite{Thaler:2004_APLa,Marcet:2006_PRB,Guo:2006_JAP,Pacuski:2007_PRB} with our findings we demonstrate that the $E_{\text{g}}$ increases linearly with the Mn content. Moreover, we are in the position to determine independently the magnitudes of $N_{0}\alpha^{\text{(app)}}$ and $N_{0}\beta^{\text{(app)}}$. By examining the determined magnitudes and signs of the exchange energies as well as the dependence $E_{\text{g}}(x)$ we show that the giant Zeeman splitting of the conduction band is strongly affected by the $s-p$ exchange, as proposed theoretically by \'Sliwa and Dietl,\cite{Sliwa:2008_PRB} whereas the shift and splitting of the valence band can be explained by the non-perturbative theory of $p-d$ hybridization, put forward by Dietl.\cite{Dietl:2008_PRB}

Our paper is organized as follows: in Sections \ref{sec:samples}, \ref{sec:experiment}, and \ref{sec:theory} we provide information on the studied samples, on the experimental methods, and on the theoretical model employed to describe the spectra, respectively. Section \ref{sec:results} is divided into two parts: in the first one the dependence of the (Ga,Mn)N band gap on Mn doping is shown and discussed, while in the second one we summarize the experimental results from which the \textit{apparent} exchange energies characterizing the giant Zeeman splittings of the bands are determined. These findings are then discussed in terms of recent theoretical models.


\section {\label{sec:samples}Samples}
The studied samples have been grown by MOVPE on sapphire substrates, as described previously.\cite{Stefanowicz:2010_PRBa} They consist of a 1~$\mu$m thick GaN buffer and the (Ga,Mn)N layer with a thickness in the range 370 -- 700~nm. Bismethylcyclopentadienyl-manganese ((CH$_3$C$_5$H$_4$)$_2$Mn) was used as a Mn precursor. The detailed structural and magnetic characterization points to the absence of any secondary phases and reveals that Mn$^{3+}$ ions are randomly distributed over cation sites.\cite{Stefanowicz:2010_PRBa,Bonanni:2010_arXiv} This is further confirmed by our reflectivity measurements in the infra-red spectral region (not shown), demonstrating the presence of absorption at 1.41~eV,\cite{Marcet:2006_PRB,Korotkov:2001_PB,Wolos:2004_PRB,Zenneck:2007_JAP} originating from internal transitions within the Mn$^{3+}$ ions. 

Parameters of the studied samples are collected in Table~\ref{tab:samplelist}, where information on precursor flow rates as well as on the Mn concentration determined by SIMS and by SQUID magnetometry,\cite{Stefanowicz:2010_PRBa} is summarized.

\begin{table}
   \begin{tabular}{|c|c|c|c|c|c|}
    \hline
    MnCP$_{2}$ & \#  & Thick.  & $x_{\mathrm{Mn}}$ & $x_{\mathrm{Mn}}$ & Mn conc. \tabularnewline
    {[}sccm] & & [nm] & (SIMS)  & (SQUID)  & (SQUID) \tabularnewline
    & &  & [\%] & [\%] & [$10^{20}$ /cm$^3$]
    \tabularnewline
    \hline
    \hline
    0  & 885  & 400  &  & $<0.014$ &  $<0.05$ \tabularnewline
    \hline
    25  & 842  & 450  & 0.073  & 0.06 & 0.3\tabularnewline
    \hline
    50  & 841  & 400  & 0.14  & 0.18 & 0.8\tabularnewline
    \hline
    100  & 844  & 400  &  & 0.18 & 0.8\tabularnewline
    \hline
    125  & 849  & 400  & 0.11  & 0.14 & 0.6\tabularnewline
    \hline
    150  & 845  & 400  & 0.16  & 0.23 & 1.0\tabularnewline
    \hline
    175  & 843  & 400  &  & 0.50 & 2.2 \tabularnewline
    \hline
    200  & 831  & 200  &  & 0.21 & 0.9\tabularnewline
    \hline
    225  & 850 & 370  & 0.25  & 0.37 & 1.6\tabularnewline
    \hline
    250  & 851  & 370  &  & 0.32 & 1.4\tabularnewline
    \hline
    275  & 854  & 400  &  & 0.37 & 1.6\tabularnewline
    \hline
    300  & 852  & 400  & 0.30  & 0.32 & 1.4\tabularnewline
    \hline
    325  & 856  & 400  &  & 0.50 & 2.2\tabularnewline
    \hline
    350  & 853  & 370  &  & 0.50 & 2.2\tabularnewline
    \hline
    375  & 857  & 400  & 0.43  & 0.57 & 2.5\tabularnewline
    \hline
    400  & 855  & 370  &  & 0.59 & 2.6\tabularnewline
    \hline
    475  & 888  & 700  &  & 0.62 & 2.7\tabularnewline
    \hline
    490  & 889 & 700 & 0.55 & 0.87 & 3.8\tabularnewline
    \hline
  \end{tabular}
   \caption{List of the samples studied in this work with the Mn precursor flow rate, the approximate thickness of the Mn-doped layer, and the Mn concentrations as determined by SIMS and SQUID. All samples have been grown with 1500 sccm NH$_3$ and 5 sccm Ga(CH$_3$)$_3$ flow rates.}
  \label{tab:samplelist}
\end{table}

\section {\label{sec:experiment}Experimental}

For the excitation of excitonic photoluminescence we employ a Kimmon He-Cd laser with the main line at 325~nm. The emitted light is focused onto the entrance slit of a Jobin Yvon Triax 550 spectrometer equipped with three gratings of 2400, 1800, 1000 groves/mm and coupled to a $1024\times 128$ pixel liquid nitrogen cooled CCD array. A 340~nm edge filter is used to block stray laser light.

The magnetoreflectivity studies are carried out in Faraday configuration for two circular light polarizations. A high pressure Xe lamp serves as light source. The beam impinging onto the sample at normal incidence is focused to a 0.2~mm diameter spot on the sample surface. The spectra are acquired by a Peltier cooled CCD camera coupled to a 2400~gr/mm grating spectrometer. The measurements are performed at temperatures between 1.6 and 100~K and magnetic fields up to 8~T.

\section{\label{sec:theory}Model description of spectra}

Symmetric Lorentzian curves are fitted to the photoluminescence spectra  in order to determine accurately the positions and intensities of particular peaks.\cite{Schubert:1986_PRB,Korona:2002_PRB} The dependence of the emission intensity on temperature obtained in this way serves for a plausible identification of the specific transitions. Further information on the origin of the transitions is obtained by examining the coupling to phonons, as well as by comparing the energy of the transitions seen in PL and reflectivity.

In order to determine the energy positions of the excitonic transitions from reflectivity spectra, the model developed by Pacuski {\em et al.}\cite{Pacuski:2008_PRL} for (Ga,Fe)N is adapted to our case. The model takes into account contributions to the (Ga,Mn)N layer and the GaN buffer dielectric functions due to the absorption by $A$, $B$, and $C$ free excitons and their excited states. As in Ref.~\onlinecite{Pacuski:2008_PRL}, the transitions to the continuum of states are modeled following Tanguy.\cite{Tanguy:1995_PRL} The square roots of the respective dielectric functions give the energy-dependent refractive indices of the (Ga,Mn)N and GaN layers. The reflectivity from the whole structure, including the substrate, is calculated $via$ the transfer matrix method taking into account Fabry-Perot-like interferences. The analytical procedure for the determination of the reflectivity spectra from samples consisting of three layers with different refractive indices is summarized in the Appendix.

Input parameters of the model are: the refractive index of the Al$_2$0$_3$ substrate taken as 1.8; the background dielectric constant $\epsilon_{0} = 5.2$ in GaN and (Ga,Mn)N; the exciton binding energies $R$$_{A}^{\ast} = 27.2$~meV, $R$$_{B}^{\ast} = 23.9$~meV, $R$$_{C}^{\ast} = 26.5$~meV, for both GaN and (Ga,Mn)N; and a damping parameter $\Gamma_{\infty} \mbox{[(Ga,Mn)N]} = 15$~meV, $\Gamma_{\infty}\mbox{[GaN]} = 8$~meV common to all excited states in (Ga,Mn)N and GaN, respectively. The free parameters of the fit are the energies, the polarizabilities, and the damping rates of the $A$, $B$ and $C$ excitons in GaN and (Ga,Mn)N at a given temperature, magnetic field, and circular polarization.

\section {\label{sec:results}Experimental results and discussion}
\subsection {\label{sec:bandgap}{(Ga,Mn)N} band gap}

\begin{figure}
\includegraphics[width=0.8\linewidth]{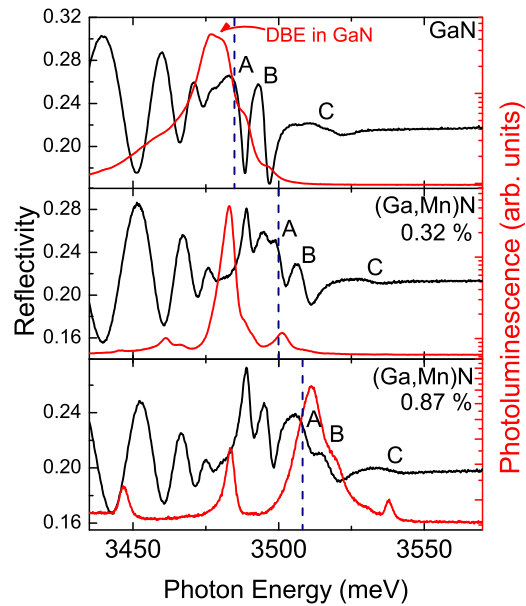}
\caption{(Color online) Left axis: Reflectivity spectra from GaN and (Ga,Mn)N with the Mn concentration $x= 0.32$\% and 0.87\% recorded at $T = 2$~K. Right axis: photoluminescence intensity acquired at $T = 10$~K for the same samples. The structure around 3.48 eV is attributed to donor bound excitons in the GaN buffer layer. The position of the (Ga,Mn)N $A$ exciton as determined from reflectivity is indicated with a dashed line in each panel. The maxima at 3.446~eV and 3.537~eV in the PL spectrum of the 0.87\% sample are assigned to 3$^{rd}$ and 4$^{th}$ order Raman scattering of the laser light (E$_{laser}$ = 3.814~eV) with LO phonon ($E_\text{{LO}}$ = 92~meV in GaN).
} \label{fig:Spectra}
\end{figure}

The experimental PL and reflectivity spectra are displayed in Fig.~\ref{fig:Spectra} for different Mn concentrations. Due to the relatively small thickness of the Mn-doped layers -- $t \approx 400$~nm -- photocarriers are generated also in the buffer layer. Neutral donor bound excitons (DBE) in the GaN buffer and free excitons in the (Ga,Mn)N layer dominate the PL spectrum. The strongest emission from the doped layer is caused by free excitons, since there are no neutral residual donors present due to their compensation by Mn.\cite{Graf:2002_APL} At the same time, scattering of free excitons by charged donors and Mn impurities may relax the $k$ conservation rules, enhancing the free exciton radiative recombination. Charged donor bound excitons, in principle competing with free excitons, are usually not observed in GaN.\cite{Monemar:2001_JPCM,Monemar:2008_pssb} As indicated in Fig.~\ref{fig:Spectra}, the PL spectra of the doped layer contain usually two ($A$ and $B$) of the three free exciton lines proper of the wurtzite structure.

The distance between the excitons in the Mn-doped layer and in the GaN buffer is used to determine the variation of the band gap with the Mn concentration $x$. Since the free exciton $A$ is not clearly resolved in the buffer layer for some samples, is such cases the DBE position is taken as a reference.  The energy difference between the exciton $A$ and DBE in the buffer, as measured for eight samples at $30$~K, is found to be $8.35\pm 0.25$~meV.

In the reflectivity spectra only free excitons are visible due to their relatively high density of states.\cite{Kornitzer:1999_pssb} The reflectivity signal of the buffer and thin (Ga,Mn)N layers is of a comparable strength. Here, the transitions of all three free $A$, $B$ and $C$ excitons in the Mn-doped layer are well resolved even for higher Mn concentration, as reported in Fig.~\ref{fig:Spectra}. The buffer layer excitons are also clearly visible, since their energies fall in the transparent region of the Mn-doped layers. Their positions do not vary with increasing Mn concentration. The lines assigned to excitons in (Ga,Mn)N are seen to undergo a shift to higher energies and to broaden with increasing Mn content.

\begin{figure}
\includegraphics[width=1\linewidth]{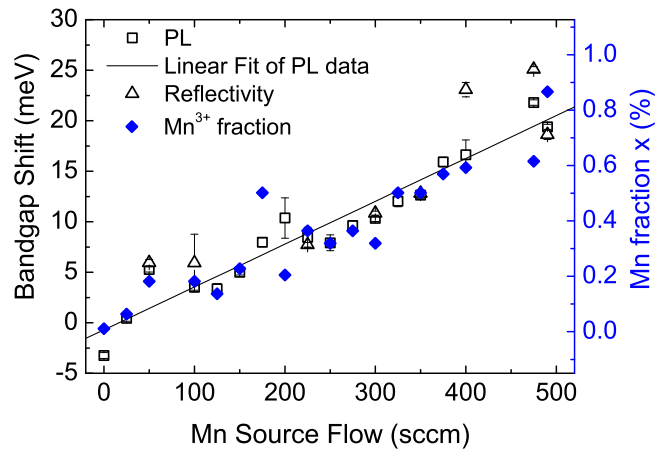}
\caption{(Color online) Relative energy position of the (Ga,Mn)N $A$ exciton with respect to GaN, as determined from PL (squares) and reflectivity (circles), plotted $vs.$ the Mn precursor flow rate. The Mn concentration determined by SQUID (diamonds) is presented on the right axis. The error bars are a combination of experimental and fitting errors. In the case of the samples with $x<0.20$\%, the energies of (Ga,Mn)N and GaN excitons overlap, resulting in an increased error bar.}
\label{fig:shift}
\end{figure}

The difference of free exciton energies between the doped layer and the buffer, as determined from the PL and reflectivity spectra according to the fitting procedure outlined in Sec.~\ref{sec:theory}, is plotted  in Fig.~\ref{fig:shift} $versus$ the Mn precursor flow rate.  The absolute Mn concentration scale is calibrated by SQUID magnetometry and the agreement between the exciton energies determined from reflectivity and photoluminescence is evidenced. In the studied Mn concentration range the band gap is seen to increase linearly with the Mn concentration $x$ according to,
\begin{equation}
 \Delta E_{\text{g}} = x (27.4\pm 2.6 )\, {\mbox{meV/\%}} -(0.89\pm 1.13)\, {\mbox{meV}}.
\end{equation}

An estimation of the expected variation of the (Ga,Mn)N energy gap with the incorporation of Mn atoms due to the modification of the interatomic distances  can be made by comparing the lattice parameters provided by synchrotron x-ray diffraction\cite{Stefanowicz:2010_PRBa} of Mn-doped and undoped samples and utilizing the suggested deformation potentials for GaN.\cite{Vurgaftman:2003_JAP,Yan:2009_APL} This would lead to a {\em decrease} of the band gap by $81 \pm 47$~meV for the sample with the Mn concentration $x= 0.87$\%.

A possible presence of strain associated with a lattice mismatch between the GaN buffer and (Ga,Mn)N layers has to be considered, since the changes observed here in the band gap are of the same order of magnitude as those expected for slightly compressively strained material.\cite{Kisielowski:1996_PRB}

X-ray diffraction space mapping of the sample with $x=0.59$~\% gave no hint of a difference in the diffraction patterns from the Mn-doped and from the undoped layer. This lets us to infere that within the experimental limit the stress is either not relaxed in the doped layer or the strain caused by stress relaxation is too low to be observed. Furthermore, the position of the buffer layer excitons does not show any trend with increasing Mn concentration, pointing to a negligible strain induced by the Mn-doped layer. A further measure of in-plain strain is represented by a shift of the Raman $E_2^{\text{(high)}}$ mode,\cite{Kisielowski:1996_PRB} but from Raman measurements on our samples with different Mn-contents (not shown) we find that all determined $E_2^{\text{(high)}}$ values are slightly scattered within 0.6~cm$^{-1}$.

By considering the reciprocal space mapping, the buffer layer exciton position in PL, the Raman spectra,  and the theoretical calculations for an upper limit of strain allowed by experimental uncertainties, we can conclude that the strain induced by the lattice mismatch of the GaN buffer and the (Ga,Mn)N doped layer cannot be solely responsible for the observed increase of the band gap. Moreover, the expected increase of the gap from Vegard's law is an order of magnitude smaller than observed.

An important source of gap variation in DMSs is the $p-d$ hybridization between valence and TM states, that produces a short-range attractive potential for holes at the TM impurities, particularly large for compounds with short bond lengths, like nitrides and oxides.\cite{Dietl:2008_PRB,Benoit:1992_PRB}  The effect of $p-d$ hybridization, if evaluated within the virtual crystal approximation (VCA), leads to a {\em decrease} of the gap by about 20~meV for the range of Mn concentrations considered here.\cite{Dietl:2008_PRB} However, if the strength of the attractive potential  increases, the VCA ceases to be valid, particularly in the strong coupling limit, where the TM ion can bind a hole. According to the generalized alloy theory,\cite{Dietl:2008_PRB,Tworzydlo:1994_PRB} which determines the effects of $p-d$ coupling in a non-perturbative way, the presence of hole bound levels renormalizes significantly the extended valence band states. In this range, the band gap is expected to {\em increase} roughly linearly with the TM concentration,\cite{Dietl:2008_PRB} as observed here. A strong hole localization, and the associated absence of itinerant holes explains also the low Curie temperatures observed even at relatively high Mn concentrations.\cite{Bonanni:2010_arXiv,Sarigiannidou:2006_PRB}

\subsection {\label{sec:magnetoref}Magneto-reflectivity of (Ga,Mn)N}

In order to further clarify the mechanisms of exchange interaction between Mn ions and band carriers in (Ga,Mn)N, reflectivity as a function of a magnetic field has been studied. The quality of the samples allows us to observe the Zeeman splitting of all three excitons, $A$, $B$, and $C$.

\begin{figure}
\includegraphics[width=0.8\linewidth]{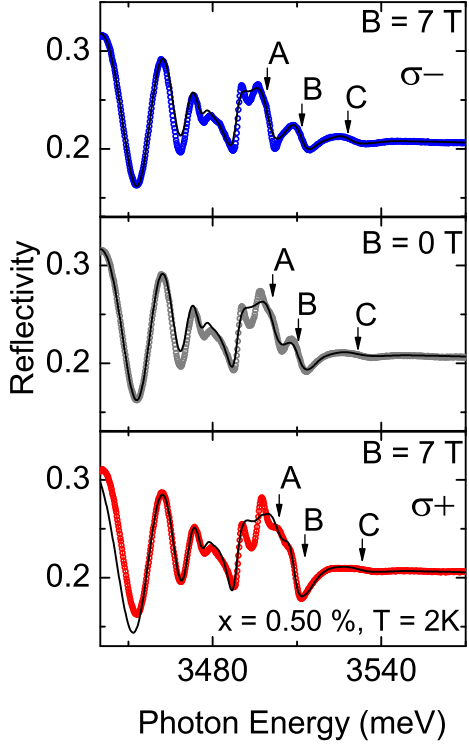}
\caption{(Color online) Reflectivity of (Ga,Mn)N (Mn content $x = 0.5$\%) in magnetic field 0~T and 7~T (Faraday configuration) at $T=2$~K. The energy positions of the $A$, $B$ and $C$ excitons at $B= 7$~T and for the $\sigma^-$ circular polarization obtained from the fit are indicated. Points - experimental data, solid lines - model.} \label{fig:magnetorefspec}
\end{figure}

In Fig.~\ref{fig:magnetorefspec} the reflectivity spectra for the sample with $x=0.50$\% recorded in the magnetic field of 0~T and 7~T for the two circular polarizations of the detected signal are reported. The transitions of $A$, $B$ as well as $C$ excitons from both GaN and (Ga,Mn)N are resolved. As expected, the contributions from the excitonic transitions in GaN and in (Ga,Mn)N are spectrally rather close and the measurement is affected by interferences resulting from multiple reflections of the light at the interfaces.

As shown in Fig.~\ref{fig:magnetorefspec}, the model employed allows us to describe properly the shape and intensity of the reflectivity in the excitonic region as well as the oscillations at lower energies originating from the interferences. It is true also in the case of the sample with the highest Mn concentration $x=0.87$\%, for which excitonic lines become strongly broadened (not shown). The determination of the excitonic shifts is not possible in the case of (Ga,Mn)N with the lowest Mn content due to spectral overlap of the transitions from the GaN buffer and the Mn-doped layer.

\begin{figure}
\includegraphics[width=0.8\linewidth]{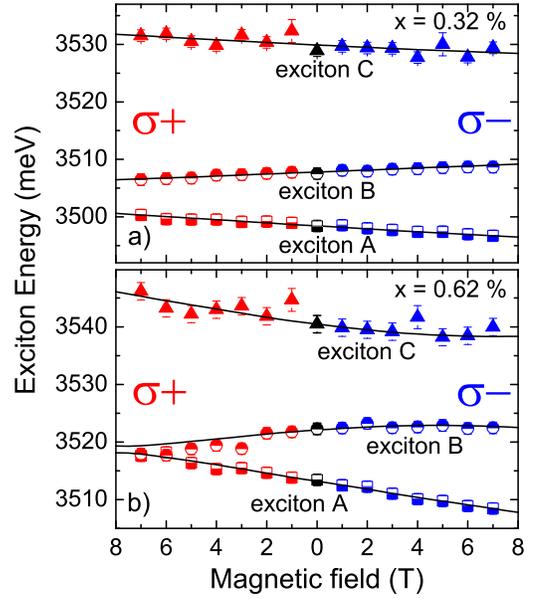}
\caption{(Color online) Exciton energies in (Ga,Mn)N (a) $x = 0.32$\%, (b) $x = 0.62$\% as a function of the magnetic field at $T = 1.8$~K. Points - experimental data, solid lines - theory.} \label{fig:fitZeeman}
\end{figure}

The identification of the characteristic excitonic transitions in (Ga,Mn)N  previously discussed has been confirmed by the observation of giant Zeeman splitting of the lines in magnetic field, as shown in Fig.~\ref{fig:fitZeeman}. The splitting has been found to persist up to $T = 100$~K (not shown).

In Fig.~\ref{fig:fitZeeman} the energy positions of the $A$, $B$, and $C$ excitons in (Ga,Mn)N, as determined from the fit for the samples with Mn concentrations of $0.32$~\% and $0.62$~\%, are presented as a function of the magnetic field. The energy difference between the positions of the $A$ and $B$ excitons increases with the magnetic field for the $\sigma^-$ polarization and decreases when the polarization is reversed. As it can be expected, the excitonic shifts are enhanced in the case of the sample with the higher Mn content. When the magnitude of the excitonic splitting increases, the shifts cease to be proportional to the magnetization because of anticrossings between exciton states, as shown in Fig.~\ref{fig:fitZeeman}(b). The anticrossing of excitons $A$ and $B$ occurs in $\sigma^+$ polarization, and is driven by an  electron-hole exchange interaction, whereas the $B$ and $C$ anticrossing results from a spin-orbit coupling.\cite{Pacuski:2007_PRB} The exciton Zeeman splittings are, therefore, not simply proportional to the magnitude of the magnetization, in contrast to DMSs with a larger separation between excitonic states.\cite{Gaj:1979_SSC}

In order to determine the constants characterizing the $s,p-d$ exchange interactions, the model of excitons in wurtzite DMSs\cite{Pacuski:2008_PRL} is fitted to the excitonic shifts in magnetic field. The model assumes an effective excitonic Hamiltonian of the form,
\begin{equation}
H = E_{0} + H_{\text{vb}} + H_{\text{Z}} + H_{\text{e-h}} + H_{\text{diam}} + H_{s,p-d},
\end{equation}
where $E_{0}$ is the band-gap energy, $H_{\text{vb}}$ describes the structure of the valence band at $k = 0$ in wurtzite semiconductors, taking into account a trigonal component of the crystal field, biaxial strain, and an anisotropic spin-orbit interaction. The term $H_{\text{Z}}$ represents a standard Zeeman excitonic Hamiltonian;  $H_{\text{e-h}}$ accounts for electron-hole Coulomb and exchange interactions within the exciton; $H_{\text{diam}}$ describes a diamagnetic shift that is quadratic in the magnetic field.

We observe a remarkable and Mn-concentration dependent shift of the excitonic states, demonstrating the significance of the $s,p-d$ exchange couplings in (Ga,Mn)N. Following the long-established approach for DMSs, we describe the giant Zeeman splitting of the free excitonic states through the $s,p-d$ exchange hamiltonian in the virtual crystal and molecular-field approximations, $H_{s,p-d} = H_{s-d} +H_{p-d}$, where
\begin{equation}
H_{s-d} = - N_{0}\alpha^{\text{(app)}}x \langle \bm{S} \rangle \textbf{s}_{e};
\end{equation}
\begin{equation}
H_{p-d} = - N_{0}\beta^{\text{(app)}}x \langle \bm{S} \rangle \textbf{s}_{h}.
\end{equation}
Here, $N_{0}\alpha^{\text{(app)}}$ and $N_{0}\beta^{\text{(app)}}$  are the \textit{apparent} exchange integrals for electrons in the conduction band and holes in the valence band, respectively, whose magnitudes and even signs may differ from the bare values according to recent theoretical suggestions;\cite{Dietl:2008_PRB,Sliwa:2008_PRB} $x$ is the Mn concentration, $\langle \bold{S} \rangle$ is the mean spin of the magnetic dopants, $\textbf{s}_{e}$ and $\textbf{s}_{h}$ are spin operators of the electron and hole, respectively.
The projection $\langle S_{z} \rangle$ of paramagnetic Mn$^{3+}$ ions is calculated as a function of temperature  and the magnetic field along the $c$ axis assuming the parallel Land\'e factor $g_{\|} = 1.91$ and the spin-orbit splitting $D = 0.27$ meV.\cite{Stefanowicz:2010_PRBa}

The values of the exchange constants $N_{0}\alpha^{\text{(app)}}$ and $N_{0}\beta^{\text{(app)}}$, the band gap energy $E_{0}$, and the valence band spin-orbit splittings $\Delta_{1}$ and  $\Delta_{2}$ constitute fitting parameters of the model. The Zeeman splitting of the excitons $A$ and $B$ is proportional to $N_{0}(\alpha^{\text{(app)}}- \beta^{\text{(app)}})$, whereas the splitting of the exciton $C$ is proportional to $N_{0}(\alpha^{\text{(app)}}+ \beta^{\text{(app)}})$. Hence, the data provide sufficient information to determine independently both exchange constants.

\begin{figure}
\includegraphics[width=0.9\linewidth]{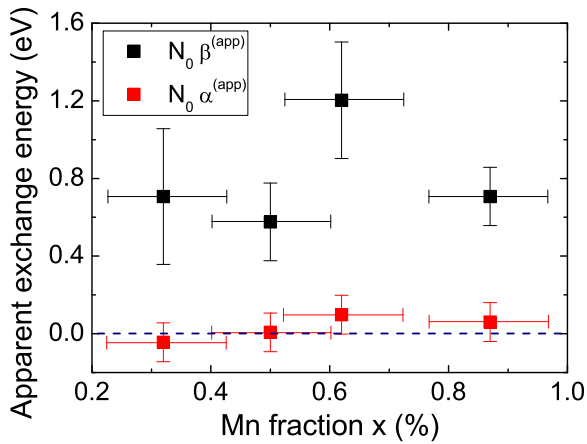}
\caption{Exchange constants $N_{0}\alpha^{\text{(app)}}$ and $N_{0}\beta^{\text{(app)}}$ for (Ga,Mn)N determined from a fit of the $A$, $B$ and $C$ excitonic shifts in magnetic field for samples with the Mn concentration from 0.32 \% to 0.87 \%.} \label{fig:exch_const}
\end{figure}

As shown in Fig.~\ref{fig:fitZeeman}, the model allows for a quantitative description of the experimental data. The fitting procedure leads to $N_{0}\alpha^{\text{(app)}} = 0.0 \pm 0.1$~eV and $N_{0}\beta^{\text{(app)}} = + 0.8 \pm 0.2$~eV (Fig.~\ref{fig:exch_const}). The magnitude of the difference $N_{0}(\alpha^{\text{(app)}}- \beta^{\text{(app)}}) = + 0.8 \pm 0.2 $~eV is in agreement with the value of $+ 1.2 \pm 0.2 $~eV determined from previous magnetooptical measurements.\cite{Pacuski:2007_PRB} At the same time, however, the values and signs of $N_{0}\alpha^{\text{(app)}}$ and $N_{0}\beta^{\text{(app)}}$ are non-standard.

Indeed, previous studies on wurtzite II-VI DMSs, particularly on (Cd,TM)S, where TM = {Cr, Mn, Fe, Co}, demonstrated that $ N_{0}\alpha = + 0.21 \pm 0.04$~eV,\cite{Twardowski:1996_PRB, Herbich:1998_PRB, Nawrocki:1987_RSSP} in agreement with the standard understanding of the origin of the $s-d$ coupling in tetrahedral DMSs. The value determined here for the exchange energy $N_{0}\alpha^{\text{(app)}}= 0.0 \pm 0.1$~eV appears to contradict this insight. A similar disagreement had been found in the case of (Ga,Mn)As with low Mn concentrations, $x \leq 0.13$\%, where according to magnetooptical studies $N_{0}\alpha^{\text{(app)}} = - 20 \pm 6$~meV.\cite{Stern:2007_PRB} The latter was explained\cite{Sliwa:2008_PRB} by noting that the Mn$^{3+}$ center consists there of five $d$ electrons and a $p$-type hole, coupled by a strong antiferromagnetic $p-d$ exchange.  For such a complex, a mutual compensation of the $s-d$ and $s-p$ interactions was found to lead to $N_{0}\alpha^{\text{(app)}} \approx 0.0$,\cite{Sliwa:2008_PRB} as observed.\cite{Stern:2007_PRB} Thus, the small value of $N_{0}\alpha^{\text{(app)}}$ revealed here can be taken as an experimental indication for the $d^5 + h$ model of the Mn$^{3+}$ center in GaN,\cite{Dietl:2002_PRB,Dietl:2008_PRB} and the presence of a sizable $s-p$ exchange interaction between conduction band electrons and holes localized by Mn acceptors.\cite{Sliwa:2008_PRB}

The above model for the Mn$^{3+}$ center in GaN is consistent with results of x-ray absorption and photoemission studies carried out for (Ga,Mn)N.\cite{Hwang:2005_PRB} Those studies implied also $N_{0}\beta = -1.6$~eV. This value is in agreement with chemical trends expected within the family of (III,Mn)V compounds\cite{Dietl:2001_PRB} but appears to challenge our results which point to $N_{0}\beta^{\text{(app)}} = +0.8 \pm 0.2$~eV. This puzzle can be resolved by the recent theory,\cite{Dietl:2008_PRB} describing the effects of the $p-d$ exchange in a non-perturbative way. As discussed in the previous subsection, the strong coupling effects are particularly relevant in the case of nitrides and oxides, where---owing to the short bond length---the $p-d$ hybridization is large. This approach demonstrates that if the potential brought about by the TM impurity is strong enough to bind a hole, a substantial renormalization of the extended states takes place. In particular, the theory anticipates an increase of the band gap on TM doping and a sign reversal of the $p-d$ exchange integral describing the giant Zeeman splitting of the valence band states, both predictions confirmed qualitatively by our findings reported here for (Ga,Mn)N and previously for (Ga,Fe)N.\cite{Pacuski:2008_PRL} However, as already noted,\cite{Pacuski:2008_PRL} in (Ga,Mn)N -- in contrast to (Ga,Fe)N -- the magnitude of $N_{0}\beta^{\text{(app)}}$ can be affected by an exchange coupling between two holes: one within the exciton and another one residing on the Mn ion. The strength of this $p-p$ exchange interaction is so-far unknown.

\subsection{Conclusions}

All three fundamental free excitons $A$, $B$, and $C$ have been observed in photoluminescence and reflectivity experiments on thoroughly characterized paramagnetic (Ga,Mn)N epilayers, in which, owing to small donor compensation, the great majority of Mn ions is in the 3+ state. The excitonic energies have been determined as a function of the magnetic field for the Mn concentration $x \leq 0.87$\%. An increase of the (Ga,Mn)N band gap with increasing Mn concentration has been demonstrated by means of PL and reflectivity studies.
Furthermore, measurements carried out in magnetic field have yielded effective values of the exchange energies $N_{0}\beta^{\text{(app)}} = +0.8 \pm 0.2$ eV and $N_{0}\alpha^{\text{(app)}} = 0.0 \pm 0.1$ eV. The determined variation of the band gap as well as the non-standard sign and magnitude of the effective exchange constants corroborate recent theoretical works on the $s,p-d$ exchange interaction for DMSs in strong coupling regime, where the TM impurity gives rise to a hole bound state.\cite{Dietl:2008_PRB,Sliwa:2008_PRB} These findings imply, in particular, the $d^5 + h$ configuration for the Mn$^{3+}$ ion in GaN. Owing to strong $p-d$ hybridization, the holes are tightly bound, and their delocalization is hampered. The corresponding absence of itinerant holes in (Ga,Mn)N explains the low Curie temperatures observed even at relatively high Mn concentrations.\cite{Bonanni:2010_arXiv,Sarigiannidou:2006_PRB}

\section*{Acknowledgments}
We acknowledge the support by the European Commission through the FunDMS Advanced Grant of the ERC within the "Ideas" 7th Framework Programme, the Austrian Fonds zur F\"{o}rderung der wissenschaftlichen Forschung-FWF (P22477, P20065 and N107-NAN), by polish NCBiR
project LIDER and the \"OAD Exchange Program Poland-Austria P13/2010. We thank Tomasz Dietl for valuable discussions.

\appendix*
\section{Description of reflectivity spectra including interferences}
We give an analytical expression to describe the reflectivity spectra for a structure consisting of three layers with refractive indices $n_1$, $n_2$ and $n_3$, respectively. The transfer matrix method\cite{Born:1959_B} is employed for an electromagnetic wave impinging onto the sample under normal incidence angle from a medium characterized by the refractive index $n_0 = 1$. Boundary conditions at the layer interfaces and the light propagation within the layers are taken into account. The thickness of the $s$-th layer ($s = 1,2$) is $d_s$, while the layer $3$ is assumed to be much thicker than the remaining two. As a consequence, the reflection from the back side of this layer is neglected.

The reflectivity $R$ for a given wavelength $\lambda$ is,

\begin{equation}
R = \left| \frac{A+B}{A-B} \right |^2,
\end{equation}
where
\begin{equation}
\begin{split}
A = n_1 n_2 p_1 (n_2 m_2 - n_3 p_2) + n_1^2 m_1 (n_2 p_2 - n_3 m_2),
\end{split}
\end{equation}
\begin{equation}
\begin{split}
B = n_2 m_1 (n_2 m_2 - n_3 p_2 ) + n_1 p_1 (n_2 p_2 - n_3 m_2 ).
\end{split}
\end{equation}

The factors $m_s$ and $p_s$ are given by,
\begin{equation}
p_s = \exp(i 2 \phi_s)+1,
\end{equation}
\begin{equation}
m_s =\exp(i 2 \phi_s)-1,
\end{equation}

where the phase factor $\phi_s$ is defined as $\phi_s = 2\pi n_s d_s/\lambda$.

\bibliography{Bibliography_JS}
\end{document}